\documentclass[prl,twocolumn,floatfix,reprint,aps,superscriptaddress]{revtex4}

\usepackage{amsfonts}
\usepackage{mathrsfs}
\usepackage{amsmath}
\usepackage{color}
\usepackage{graphicx}
\usepackage{bm}
\usepackage{amssymb}
\usepackage{xspace}
\usepackage{epstopdf}
\usepackage{dcolumn}
\usepackage{longtable}
\usepackage{multirow}
\usepackage[colorlinks=true, letterpaper=true, pdfstartview=FitV, linkcolor=blue, citecolor=blue, urlcolor=blue]{hyperref}

\begin{document}

\title{Hourglass Dirac Chain Metal in Rhenium Dioxide}

\author{Shan-Shan Wang}
\affiliation{Research Laboratory for Quantum Materials, Singapore University of Technology and Design, Singapore 487372, Singapore}

\author{Ying Liu}
\affiliation{Research Laboratory for Quantum Materials, Singapore University of Technology and Design, Singapore 487372, Singapore}

\author{Zhi-Ming Yu}
\affiliation{Research Laboratory for Quantum Materials, Singapore University of Technology and Design, Singapore 487372, Singapore}

\author{Xian-Lei Sheng}
\email{xlsheng@buaa.edu.cn}
\affiliation{Research Laboratory for Quantum Materials, Singapore University of Technology and Design, Singapore 487372, Singapore}
\affiliation{Department of Applied Physics, Key Laboratory of Micro-nano Measurement-Manipulation and Physics (Ministry of Education), Beihang University, Beijing 100191, China}

\author{Shengyuan A. Yang}
\email{shengyuan\_yang@sutd.edu.sg}
\affiliation{Research Laboratory for Quantum Materials, Singapore University of Technology and Design, Singapore 487372, Singapore}

\begin{abstract}
Nonsymmorphic symmetries, which involve fractional lattice translations in crystalline materials, can generate exotic types of fermionic excitations that are robust against spin-orbit coupling. Here we report on a hourglass-type dispersion in the bulk of three-dimensional rhenium dioxide crystals, as dictated by its nonsymmorphic symmetries. Due to time reversal and inversion symmetries, each band has an additional two-fold degeneracy, making the neck crossing-point of the hourglass four-fold degenerate. Remarkably, close to the Fermi level, the neck crossing-point traces out a Dirac chain---a chain of connected four-fold-degenerate Dirac loops---in the momentum space. The symmetry protection, the transformation under symmetry-breaking, and the associated topological surface states of the hourglass Dirac chain are discussed.
\end{abstract}


\maketitle

Topological metals or semimetals, which host robust fermionic excitations around protected band-crossing points, have been a focus of current research. For example, Weyl and Dirac semimetals possess two- and four-fold degenerate isolated band-crossing points close to the Fermi level, around which the quasiparticles resemble the relativistic Weyl and Dirac fermions~\cite{Wan2011,Murakami2007,Burkov2011,Volovik2003,Young2012,Wang2012b,Wang2013b,Zhao2013c,Yang2014a,Weng2015,Huang2015,Liu2014c,Borisenko2014,Lv2015,Xu2015a}. Under certain symmorphic symmetry operations such as mirror or inversion, the crossing points may also form one-dimensional (1D) nodal loops~\cite{Weng2015c,Yang2014c,Mullen2015,Yu2015,Kim2015a,Chen2015,Xie2015,Fang2015,Chan2016,Li2016,Bian2016,Schoop2016,HXu,Yu2017}, but such loops are usually vulnerable against spin-orbit coupling (SOC) and can be removed without altering the symmetry, hence they are termed as accidental nodal loops. Recently, it was realized that nonsymmorphic symmetries, which involve fractional lattice translations, could play a key role in stabilizing the band-crossing points~\cite{Parameswaran2013,Steinberg2014,Young2015a,Watanabe2016,Bradlyn2016,Liang2016,Wieder2016,Zhao2016a,RChen}. They have two important effects. First, the degeneracies enabled by nonsymmorphic symmetries could be robust against SOC. Particularly, spin-orbit nodal loops with two- or even four-fold degeneracy have been theoretically proposed~\cite{Fang2016,Chen2016b,Yang2017,Furusaki,Takahashi}. Second, nonsymmorphic symmetries may entangle multiple bands together, so that the resulting crossing points are unavoidable and entirely dictated by the crystalline symmetry. Such band-crossing points are thus referred to as essential. For example, it was found that bands are entangled into groups of four and form hourglass-shaped dispersion on the 2D surface of nonsymmorphic insulators KHg$X$ ($X=$As, Sb, Bi)~\cite{Wang2016a,Ma2015}. Theoretical modeling suggested that such hourglass fermions may also exist in the bulk of 3D crystals~\cite{Wang2017}, and interestingly, Bzdu\u{s}ek \emph{et al.}~\cite{Bzdusek2016} showed that the neck point of the hourglass may trace out a nodal chain of two-fold-degenerate Weyl loops, when multiple nonsymmorphic operations are present.

Although the essential band-crossings are solely determined by the space group for which theoretical analysis has offered valuable guidelines, the search for realistic materials that exhibit them at low energy is still challenging. This is because the bands in real materials typically have complicated 3D dispersions, such that the crossing point that we are chasing may be far away from the Fermi energy. The situation could be even worse for nodal loops, since the points on the loop are not guaranteed to have the same energy, there might be large energy variation around the loop. So far, the proposed nonsymmorphic topological metals are still limited, therefore, it is urgent to discover more suitable candidate materials to expedite studies of their intriguing properties.

\begin{figure}[t!]
\includegraphics[width=9.2cm]{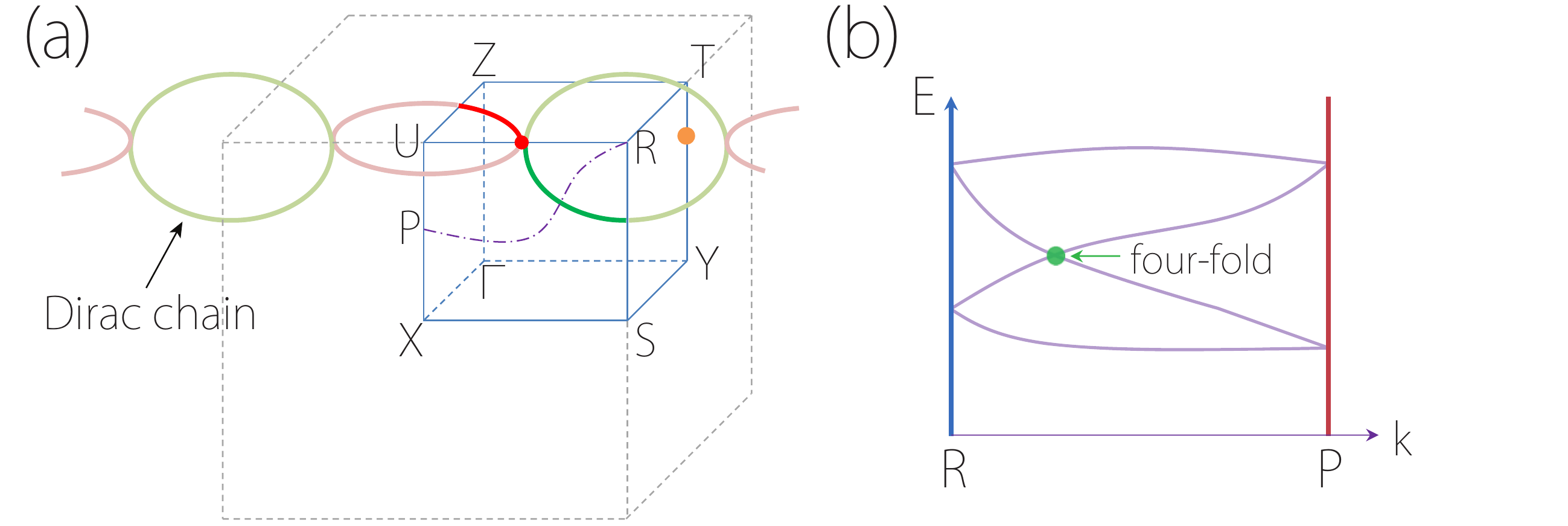}
\caption{(a) Schematic figure showing Dirac chain in ReO$_2$, which consists of one (red) loop in $k_z=\pi$ plane and one (green) loop in $k_x=\pi$ plane. There is another isolated Dirac point (orange dot) on T-Y. These crossings are four-fold degenerate and correspond to the neck crossing-point of the hourglass-type dispersion. For example, (b) shows the schematic band dispersion along a path on the $k_x=\pi$ plane connecting R and P (an arbitrary point on U-X). Each band is two-fold degenerate, and the neck point (green dot) is four-fold degenerate.}
\label{fig1}
\end{figure}

In this paper, based on first-principles calculations and symmetry analysis, we report on hourglass-type essential band-crossings in an existing material---ReO$_2$. We show that due to time reversal ($\mathcal{T}$) and inversion ($\mathcal{P}$) symmetries, the hourglass here is actually doubled, and the neck crossing-point here becomes a Dirac point with four-fold degeneracy [Fig.~\ref{fig1}(b)]. Remarkably, close to Fermi level, the neck point traces out a Dirac chain---a chain of connected (four-fold-degenerate) Dirac loops---in the momentum space, as schematically shown in Fig.~\ref{fig1}(a), hence the state may be dubbed as an hourglass Dirac chain metal. This chain is essential, robust against SOC, and dictated by two orthogonal glide mirror planes combined with time-reversal and inversion symmetries. In addition, there is another pair of bulk hourglass Dirac points on a symmetry line [see Fig.~\ref{fig1}(a)]. We clarify the protection of these exotic band-crossings, and discuss their transformations under symmetry-breaking as well as the associated topological surface states. Our findings provide an exciting platform for explore the novel topological fermions from nonsymmorphic symmetries.

\begin{figure}[t!]
\includegraphics[width=9.2cm]{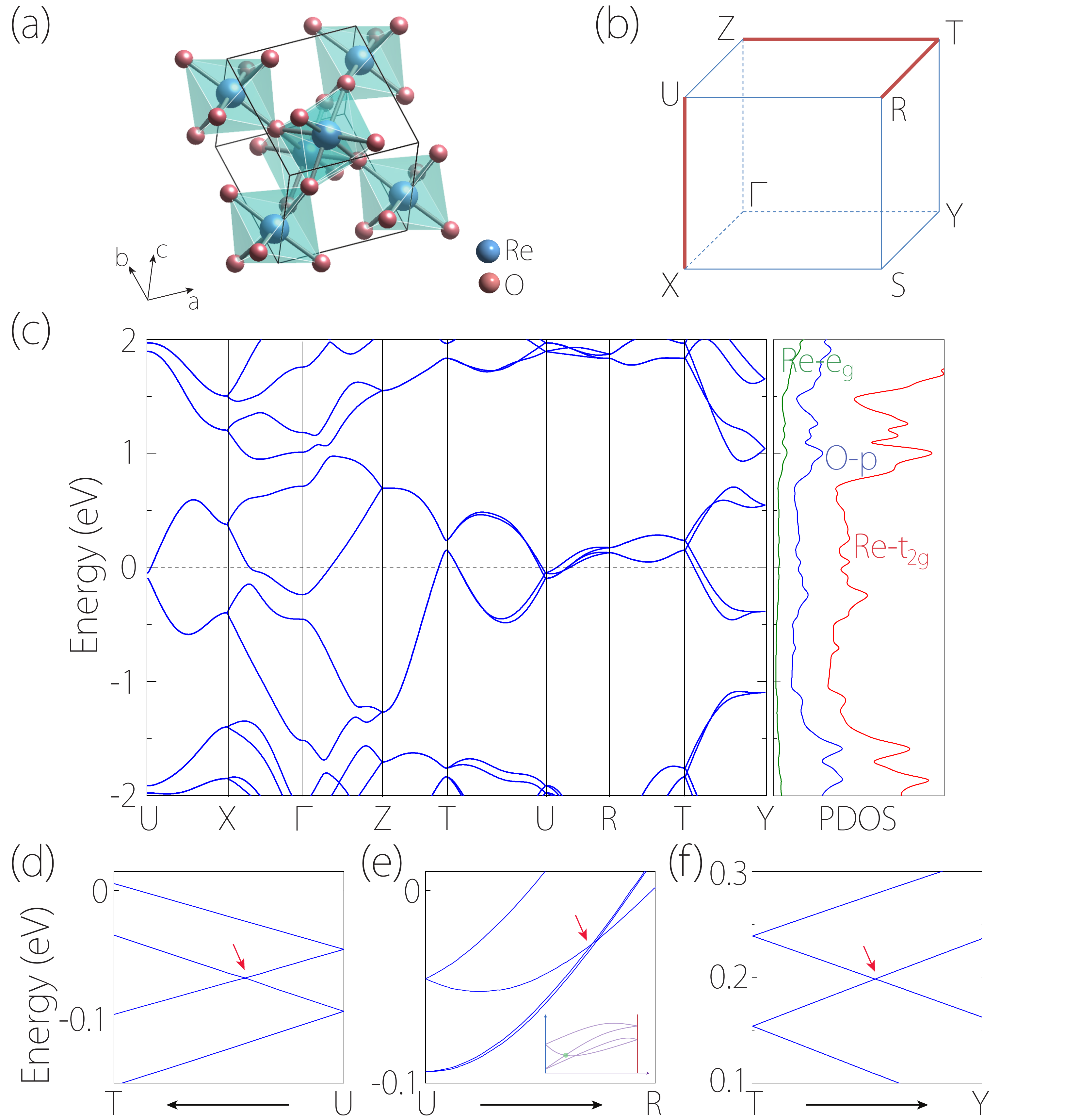}
\caption{(a) Crystal structure of ReO$_2$. (b) 1/8 Brillouin zone. The red lines indicate the paths where bands are four-fold degenerate. (c) Electronic band structure along with PDOS. (d-f) show the enlarged band structure around the four-fold-degenerate neck-crossing points. The hourglass dispersion along U-R is somewhat distorted, as schematically shown in the inset of (e).}
\label{fig2}
\end{figure}

Single crystal ReO$_2$ is observed with two structures denoted as $\alpha$ and $\beta$~\cite{Magneli1957,Colaitis1972}. $\beta$-ReO$_2$ is energetically more stable, and is found to be a stable paramagnetic metal at ambient conditions~\cite{Goodenough1965}. Hence we focus on $\beta$-ReO$_2$ here. It adopts the PbO$_2$-type orthorhombic crystal structure with space group No. 60 ($Pbcn$)~\cite{Magneli1957}. As shown in Fig.~\ref{fig2}(a), the structure is characterized by zigzag chains of Re atoms running along the $c$-axis, and each Re atom is contained in a slightly distorted octahedron of six surrounding O atoms. The space group of the structure may be generated by the following symmetry operations that will be important in our discussion: the inversion $\mathcal{P}$, and two glide mirror planes involving half lattice translations $\widetilde{\mathcal{M}}_x:(x,y,z)\rightarrow(-x+\frac{1}{2},y+\frac{1}{2},z)$ and $\widetilde{\mathcal{M}}_z:(x,y,z)\rightarrow(x+\frac{1}{2},y+\frac{1}{2},-z+\frac{1}{2})$. Here the tilde above a symbol indicates that it is a nonsymmorphic symmetry. One also notes that combining all three operations leads to a third glide mirror $\widetilde{\mathcal{M}}_y:(x,y,z)\rightarrow(x,-y,z+\frac{1}{2})$.

We performed first-principles calculations based on the density functional theory (DFT). SOC was included, and possible correlation effect of Re($5d$) orbitals was tested. The calculation details are in the Supplemental Material~\cite{supp}. The experimental values of the lattice parameters ($a=4.809$ \AA, $b=5.643$ \AA, $c=4.601$ \AA)~\cite{Magneli1957} were used in the calculation.

In octahedral crystal field, Re($5d$) orbitals are split into $t_{2g}$ and $e_g$ groups, with the latter at higher energy. For Re$^{4+}$ with 3 valence electrons, the Re-$t_{2g}$ orbitals will be half-filled, resulting in a metallic state. Figure~\ref{fig2}(c) shows the calculated band structure of ReO$_2$ along with the projected density of states (PDOS). Indeed, one observes a metallic phase with fairly dispersive bands around Fermi level, and the low-energy states are dominated by the Re-$t_{2g}$ orbitals. Understanding that each band is at least two-fold degenerate due to the presence of $\mathcal{T}$ and $\mathcal{P}$, two interesting type of band features can be observed from Fig.~\ref{fig2}(c): (i) all bands are four-fold degenerate along U-X, Z-T, and T-R; and (ii) hourglass-shaped dispersions appear on T-U, U-R, and T-Y. The neck point of the hourglass is a crossing-point with four-fold degeneracy.

Let's first investigate feature (i) regarding the four-fold degeneracy along the three high-symmetry lines. Consider the U-X line at $k_x=\pi$ and $k_y=0$ (in unit of the inverse of respective lattice parameter). It is an invariant subspace of $\widetilde{\mathcal{M}}_x$, so each Bloch state $|u\rangle$ there can be chosen as an eigenstate of $\widetilde{\mathcal{M}}_x$. Since
\begin{equation}
(\widetilde{\mathcal{M}}_x)^2=T_{010}\overline{E}=-e^{-ik_y},
\end{equation}
the $\widetilde{\mathcal{M}}_x$ eigenvalue $g_x$ must be $\pm i$ on U-X. Here $T_{010}$ denotes the translation along $y$ by one unit cell, and $\overline{E}$ is the $2\pi$ spin rotation. The commutation relation between $\widetilde{\mathcal{M}}_x$ and $\mathcal{P}$ given by
\begin{equation}\label{Eq2}
\widetilde{\mathcal{M}}_x\mathcal{P}=T_{110}\mathcal{P}\widetilde{\mathcal{M}}_x
\end{equation}
means that $\{\widetilde{\mathcal{M}}_x,\mathcal{P}\}=0$ on U-X. Consequently, each state $|u\rangle$ and its Kramers-degenerate partner $\mathcal{PT}|u\rangle$ must share the same $\widetilde{\mathcal{M}}_x$ eigenvalue. For example, assume $|u\rangle$ has $g_x=+i$ (denoted as $|+i\rangle$), then
\begin{equation}\label{Eq3}
\widetilde{\mathcal{M}}_x(\mathcal{PT}|+i\rangle)=-\mathcal{PT}(+i)|+i\rangle=i(\mathcal{PT}|+i\rangle),
\end{equation}
where in the second step we used the fact that $\mathcal{T}$ is an anti-unitary operator. Same result holds for a state with $g_x=-i$. On the other hand, U-X is invariant under another anti-unitary symmetry $\widetilde{\mathcal{M}}_z\mathcal{T}$, which also generates a Kramers-like degeneracy since $(\widetilde{\mathcal{M}}_z\mathcal{T})^2=-1$ on U-X. Note that
\begin{equation}\label{Eq4}
\widetilde{\mathcal{M}}_x\widetilde{\mathcal{M}}_z=-T_{\bar{1}00}\widetilde{\mathcal{M}}_z\widetilde{\mathcal{M}}_x,
\end{equation}
where the minus sign is due to the anti-commutativity between two spin rotations, i.e., $\{\sigma_x,\sigma_z\}=0$, so that $[\widetilde{\mathcal{M}}_x,\widetilde{\mathcal{M}}_z]=0$ on U-X. Following similar derivation in Eq.~(\ref{Eq3}), one finds that $|u\rangle$ and
$\widetilde{\mathcal{M}}_z\mathcal{T}|u\rangle$ have opposite $g_x$. Thus, the four states, $\{|u\rangle, \mathcal{PT}|u\rangle, \widetilde{\mathcal{M}}_z\mathcal{T}|u\rangle, \mathcal{P}\widetilde{\mathcal{M}}_z|u\rangle\}$ at the same $k$-point on U-X must be linearly independent and degenerate with the same energy. The four-fold degeneracy along Z-T and T-R can also be derived in a similar way~\cite{supp}.

\begin{figure}[t!]
\includegraphics[width=9.2cm]{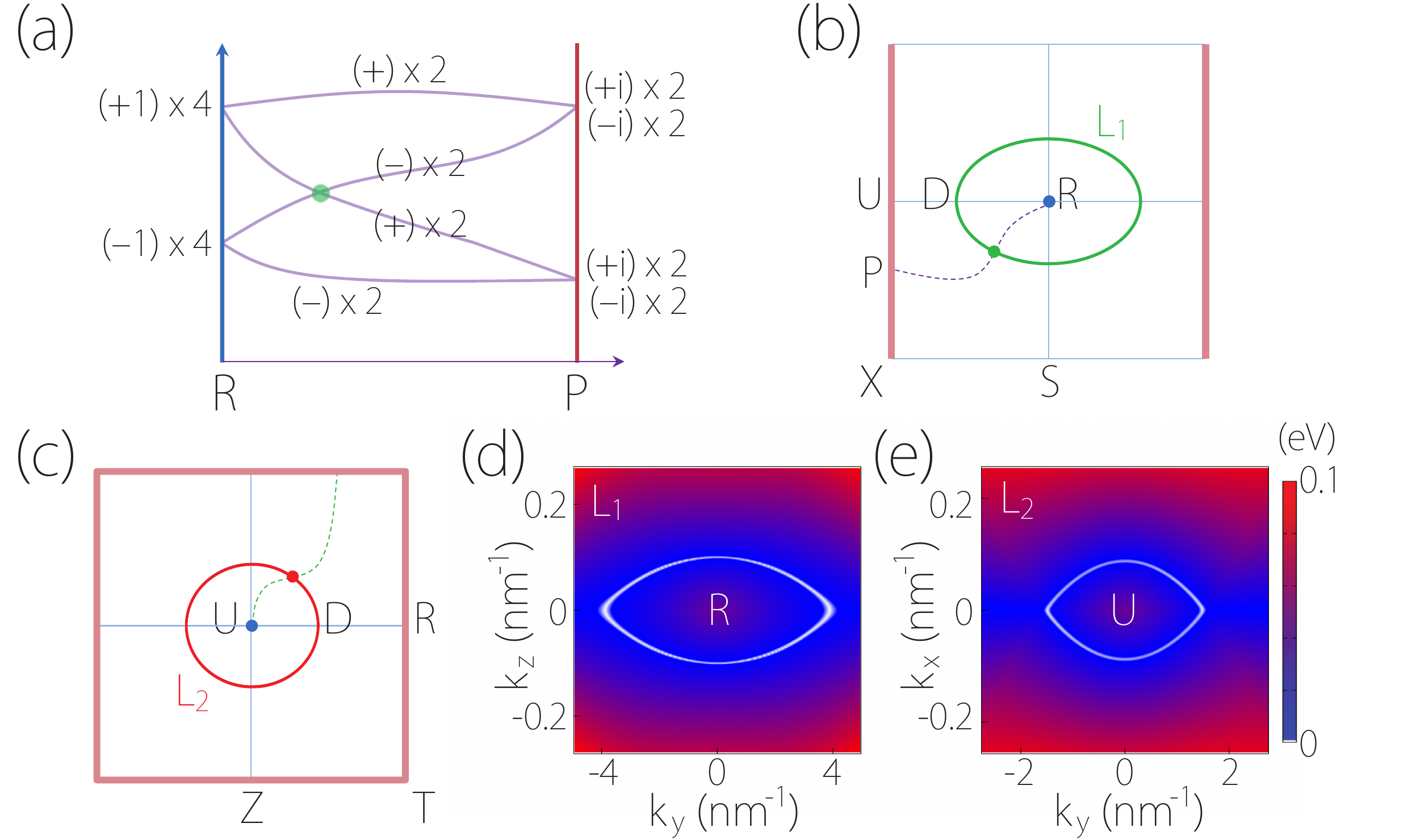}
\caption{(a) Schematic figure of hourglass dispersion along a path on $k_x=\pi$ plane connecting R to any point P on U-X (including U). The labels indicate the $\widetilde{\mathcal{M}}_x$ eigenvalues. Partner switching between two quartets leads to the four-fold-degenerate crossing point (green dot). (b,c) Such crossing traces out Dirac loop (b) $L_1$ on $k_x=\pi$ plane, and also (c) $L_2$ on $k_z=\pi$ plane. The red-colored boundaries are the lines with four-fold band degeneracy. (d,e) Shape of two Dirac loops obtained from DFT. The color-map indicates the local gap between two crossing bands. }
\label{fig3}
\end{figure}

Next, we turn to feature (ii) regarding the hourglass dispersion. Consider the line U-R. It is invariant under both $\widetilde{\mathcal{M}}_x$ and $\widetilde{\mathcal{M}}_z$. From Eq.~(\ref{Eq4}), $[\widetilde{\mathcal{M}}_x,\widetilde{\mathcal{M}}_z]=0$ on U-R, so each state $|u\rangle$ there can be chosen as simultaneous eigenstate of both operators, with eigenvalues $(g_x,g_z)=(\pm i,\pm 1)e^{-ik_y/2}$. Using the commutation relation
in (\ref{Eq2}) and $\widetilde{\mathcal{M}}_z\mathcal{P}=T_{111}\mathcal{P}\widetilde{\mathcal{M}}_z$, one finds that
\begin{equation}\label{Eq5}
(\widetilde{\mathcal{M}}_x,\widetilde{\mathcal{M}}_z)\mathcal{PT}|g_x,g_z\rangle=(g_x,g_z)\mathcal{PT}|g_x,g_z\rangle,
\end{equation}
so the Kramers pair $|u\rangle$ and $\mathcal{PT}|u\rangle$ at any $k$-point on U-R share the same $(g_x,g_z)$ eigenvalues. In addition, points R and U are time-reversal invariant momenta. At R $=(\pi,\pi,\pi)$, $(g_x,g_z)=(\pm 1,\pm i)$, hence if $|u\rangle$ has eigenvalues $(g_x,g_z)$, its Kramers partner $\mathcal{T}|u\rangle$ must have $(g_x,-g_z)$. Similarly, at U $=(\pi,0,\pi)$, since $(g_x,g_z)=(\pm i,\pm 1)$, $\mathcal{T}|u\rangle$ must have eigenvalues $(-g_x,g_z)$ if $|u\rangle$ has $(g_x,g_z)$.

Focusing on the eigenvalue $g_x$, the analysis shows that the four states in the degenerate quartet (may be chosen as $\{|u\rangle, \mathcal{T}|u\rangle, \mathcal{P}|u\rangle, \mathcal{PT}|u\rangle\}$) at R all have the same $g_x$ ($+1$ or $-1$); whereas at point U, they consist of two states with $g_x=+i$ and two other states with $g_x=-i$. Hence there has to be a switch of partners between two quartets along U-R, during which the eight bands must be entangled to form the hourglass-type dispersion. The situation is schematically shown in Fig.~\ref{fig3}(a).
It is important to note that the four-fold-degenerate neck crossing-point (denoted as D on U-R) is protected because the two crossing doubly-degenerate bands have opposite $g_x$ [with each degenerate pair sharing the same $g_x$, as shown in Eq.~(\ref{Eq5}) and illustrated in Fig.~\ref{fig3}(a)].

Furthermore, since the whole $k_x=\pi$ plane is invariant under $\widetilde{\mathcal{M}}_x$, $g_x$ is well defined for any state on this plane. Hence the above argument applies to \emph{any} path lying on the $k_x=\pi$ plane and connecting points U and R, which should feature an hourglass spectrum with four-fold-degenerate crossing-point in between.
The crossing-point must trace out a closed Dirac loop $L_1$ on this plane, as indicated in Fig.~\ref{fig3}(b). One also notes that not only U, actually any point P on U-X has four-fold degeneracy with two $g_x=+i$ and two $g_x=-i$, as we analyzed before. Thus hourglass pattern is guaranteed to appear on any path connecting R to an arbitrary point on U-X [Fig.~\ref{fig3}(b)].

Similar analysis as in the last two paragraphs applies to the $k_z=\pi$ plane, with the role played by $\widetilde{\mathcal{M}}_x$ replaced by $\widetilde{\mathcal{M}}_z$. It shows that hourglass pattern appears on any path connecting U to an arbitrary point on Z-T or T-R, and the neck point of the hourglass traces out a second Dirac loop $L_2$, as illustrated in Fig.~\ref{fig3}(c). Interestingly, $L_1$ and $L_2$ are orthogonal to each other, and they touch at the point D on the U-R line. Thus they constitute a Dirac chain in the momentum space, as shown in Fig.~\ref{fig1}(a).

Figure~\ref{fig3}(d,e) shows the locations of the Dirac loops obtained from DFT calculations, which are consistent with our symmetry analysis. The chain is close to the Fermi level and has small energy variation ($<0.2$ eV). We stress that the presence of such band-crossing pattern is solely determined by the space group (plus $\mathcal{T}$). However, whether such crossings could manifest around Fermi level and have relatively small energy variation will depend on the specific material.

Up to now, one may wonder whether there exists a third loop on the $k_y=\pi$ plane, given that $\widetilde{\mathcal{M}}_y$ is also a symmetry. It turns out not to be the case. Consider any state $|g_y\rangle$ on $k_y=\pi$ plane with $\widetilde{\mathcal{M}}_y$ eigenvalue $g_y$, one can show that
\begin{equation}
\widetilde{\mathcal{M}}_y(\mathcal{PT}|g_y\rangle)=-g_y(\mathcal{PT}|g_y\rangle).
\end{equation}
Thus each Kramers pair $|u\rangle$ and $\mathcal{PT}|u\rangle$ have \emph{opposite} $g_y$, which is in contrast with Eq.~(\ref{Eq5}) for the other two planes. As a result, $\widetilde{\mathcal{M}}_y$ can no longer protect the neck crossing-point, since each doubly-degenerate band have both  $\widetilde{\mathcal{M}}_y$ parities and two such bands would generally hybridize to open a gap. Thus a third Dirac loop on the $k_y=\pi$ plane does not appear. This is indeed confirmed by our DFT result. Nevertheless, symmetry does dictates hourglass dispersion with an isolated Dirac point on T-Y [see Fig.~\ref{fig2}(f)], due to the presence of additional $\widetilde{\mathcal{M}}_x$ symmetry on this line [Fig.~\ref{fig1}(a)]~\cite{supp}.

\begin{figure}[t!]
\includegraphics[width=9cm]{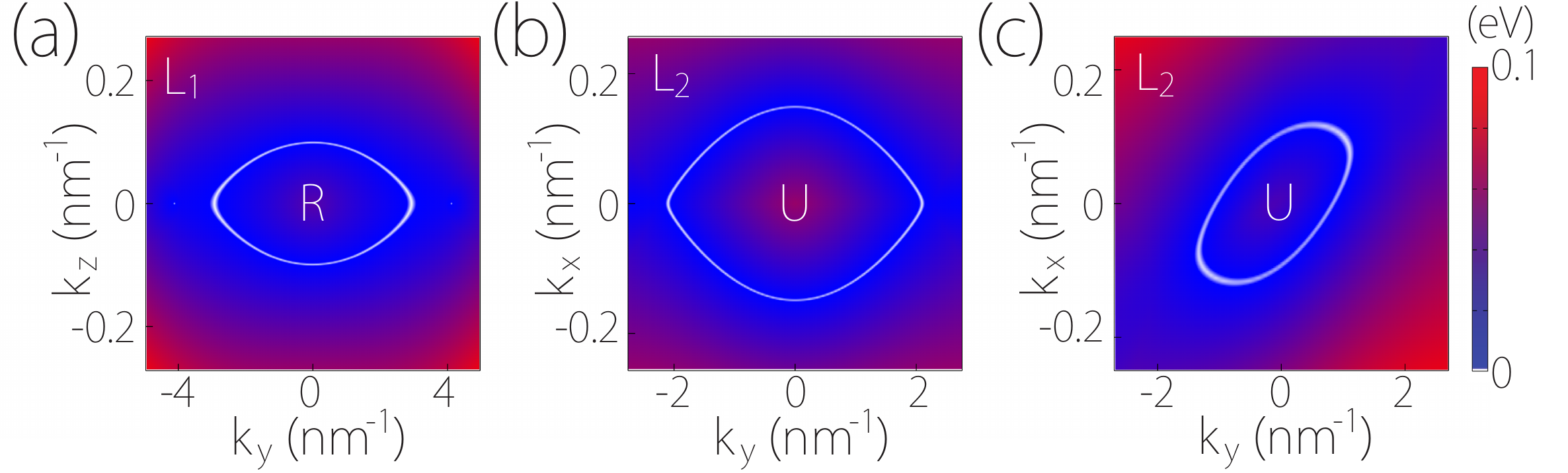}
\caption{(a,b) Dirac chain is maintained under lattice distortion that preserves the symmetry. The figures show the two loops when lattice parameters are increased by $5\%$. (c) By changing the angle between $a$ and $b$ axis (here to $80^\circ$), symmetry is reduced and only $L_2$ loop is preserved. }
\label{fig4}
\end{figure}

The hourglass dispersion and the Dirac chain are dictated by symmetry. They must be kept as long as the space group symmetry is maintained. In Fig.~\ref{fig4}(a,b), we demonstrate that when we distort the crystal lattice while maintaining the symmetry, the shape and the size of the chain can change, but it cannot be destroyed. In contrast, if we break the symmetry, e.g., by varying the angle between $a$ and $b$ axis away from 90$^\circ$ (corresponding to some shear strain) to change the lattice from orthorhombic to monoclinic, the chain will lose (part of) its protection. In this case, the distortion breaks $\widetilde{\mathcal{M}}_x$ but still preserves $\widetilde{\mathcal{M}}_z$ and $\mathcal{P}$, thus the Dirac loop on the $k_z=\pi$ plane is still protected [Fig.~\ref{fig4}(c)], whereas the loop on the $k_x=\pi$ plane and the Dirac point on T-Y are removed. These are confirmed by the DFT calculation.

\begin{figure}[t!]
\includegraphics[width=9cm]{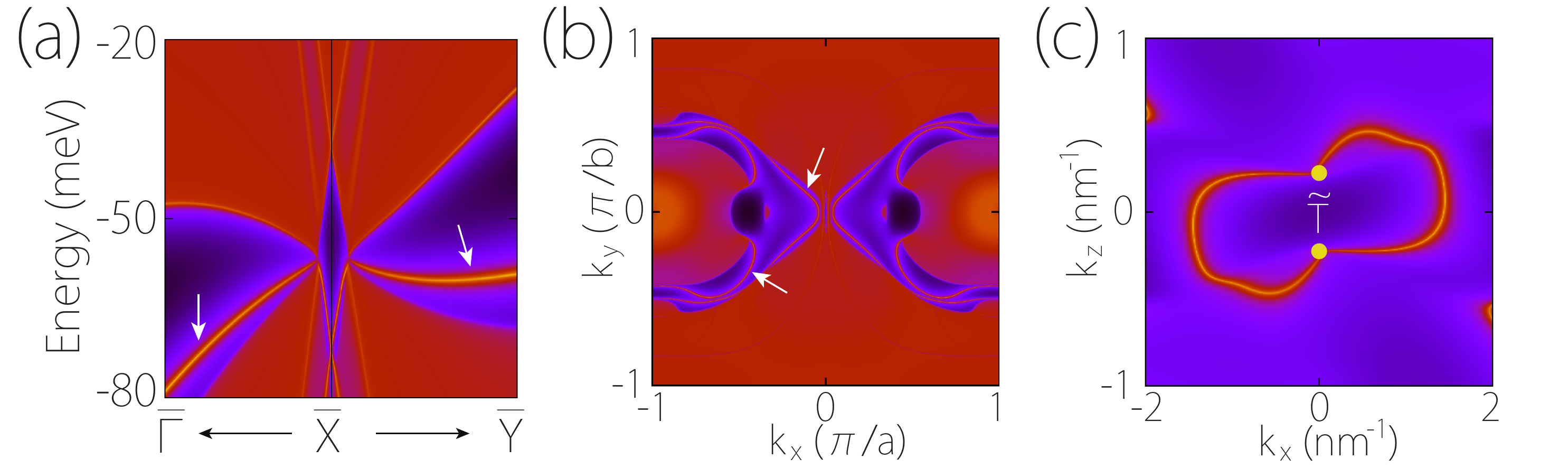}
\caption{(a) Projected spectrum on (001) surface, and (b) the corresponding constant energy slice at -60 meV. The arrows indicate the drumhead-like surface states. (c) Surface Fermi arcs on (010) surface connecting the surface-projections of the Dirac points on T-Y (marked by the orange dots).}
\label{fig5}
\end{figure}

Nodal loops could feature topological drumhead-like surface states~\cite{Weng2015c}. We find similar phenomena for the Dirac chain here. For example, on the (001) surface, the projected loop $L_2$ is centered around $\overline{\text{X}}$ point, around which one indeed observes drumhead-like surface states emanating from the projected bulk band-crossing point [Fig.~\ref{fig5}(a,b)]. The pair of isolated Dirac points on T-Y also generates surface Fermi arcs. As shown in Fig.~\ref{fig5}(c) for the (010) surface, the arcs connect the surface-projections of the bulk Dirac points, similar to the Dirac semimetals Na$_3$Bi and Cd$_3$As$_2$~\cite{Wang2012b,Wang2013b}.

A few remarks are in order before closing. First, to our knowledge, this is the first time that an hourglass Dirac chain metal is reported for a real material. The Dirac chain revealed in ReO$_2$ represents an essential band-crossing: it is robust against SOC and dictated by the crystalline symmetry. Compared with the Weyl chain studied in Ref.~\onlinecite{Bzdusek2016}, the Dirac chain here acquires additional degeneracy due to the presence of an inversion center. This important difference poses more stringent condition regarding the symmetry protection. Indeed, if there were no such degeneracy in the current case, the (missing) loop on the $k_y=\pi$ plane would be well protected.

The Dirac chain, the hourglass dispersion, and the surface states are close to the Fermi level. They could be detected in spectroscopic experiment such as ARPES~\cite{Ma2015}. Since a nodal loop has intrinsic anisotropy, the magnetic response could be very different for different magnetic field orientations. It has been shown that for essential nodal loops, there could also be pronounced anomaly in the longitudinal magnetotransport when the fields are aligned in the loop plane~\cite{Bzdusek2016}.

Finally, it has been argued that drumhead surface states could lead to huge surface density of states, which may provide a route towards high-temperature superconductivity~\cite{Heikkilae2011}. In this aspect, nodal chain metals could be ideal since the orthogonal loops dictates the presence of drumhead surface states on multiple surfaces. The size of the chain and hence the surface density of states can be tuned, e.g., by strain, which could offer an additional control of the possible phase transition.



\bibliography{HG_DChain_ref}

\begin{thebibliography}{52}
\expandafter\ifx\csname natexlab\endcsname\relax\def\natexlab#1{#1}\fi
\expandafter\ifx\csname bibnamefont\endcsname\relax
  \def\bibnamefont#1{#1}\fi
\expandafter\ifx\csname bibfnamefont\endcsname\relax
  \def\bibfnamefont#1{#1}\fi
\expandafter\ifx\csname citenamefont\endcsname\relax
  \def\citenamefont#1{#1}\fi
\expandafter\ifx\csname url\endcsname\relax
  \def\url#1{\texttt{#1}}\fi
\expandafter\ifx\csname urlprefix\endcsname\relax\def\urlprefix{URL }\fi
\providecommand{\bibinfo}[2]{#2}
\providecommand{\eprint}[2][]{\url{#2}}

\bibitem[{\citenamefont{Wan et~al.}(2011)\citenamefont{Wan, Turner, Vishwanath,
  and Savrasov}}]{Wan2011}
\bibinfo{author}{\bibfnamefont{X.}~\bibnamefont{Wan}},
  \bibinfo{author}{\bibfnamefont{A.~M.} \bibnamefont{Turner}},
  \bibinfo{author}{\bibfnamefont{A.}~\bibnamefont{Vishwanath}},
  \bibnamefont{and} \bibinfo{author}{\bibfnamefont{S.~Y.}
  \bibnamefont{Savrasov}}, \bibinfo{journal}{Phys. Rev. B}
  \textbf{\bibinfo{volume}{83}}, \bibinfo{pages}{205101}
  (\bibinfo{year}{2011}).

\bibitem[{\citenamefont{Murakami}(2007)}]{Murakami2007}
\bibinfo{author}{\bibfnamefont{S.}~\bibnamefont{Murakami}},
  \bibinfo{journal}{New Journal of Physics} \textbf{\bibinfo{volume}{9}},
  \bibinfo{pages}{356} (\bibinfo{year}{2007}).

\bibitem[{\citenamefont{Burkov and Balents}(2011)}]{Burkov2011}
\bibinfo{author}{\bibfnamefont{A.~A.} \bibnamefont{Burkov}} \bibnamefont{and}
  \bibinfo{author}{\bibfnamefont{L.}~\bibnamefont{Balents}},
  \bibinfo{journal}{Phys. Rev. Lett.} \textbf{\bibinfo{volume}{107}},
  \bibinfo{pages}{127205} (\bibinfo{year}{2011}).

\bibitem[{\citenamefont{Volovik}(2003)}]{Volovik2003}
\bibinfo{author}{\bibfnamefont{G.~E.} \bibnamefont{Volovik}},
  \emph{\bibinfo{title}{The Universe in a Helium Droplet}}
  (\bibinfo{publisher}{Clarendon Press}, \bibinfo{address}{Oxford},
  \bibinfo{year}{2003}).

\bibitem[{\citenamefont{Young et~al.}(2012)\citenamefont{Young, Zaheer, Teo,
  Kane, Mele, and Rappe}}]{Young2012}
\bibinfo{author}{\bibfnamefont{S.~M.} \bibnamefont{Young}},
  \bibinfo{author}{\bibfnamefont{S.}~\bibnamefont{Zaheer}},
  \bibinfo{author}{\bibfnamefont{J.~C.~Y.} \bibnamefont{Teo}},
  \bibinfo{author}{\bibfnamefont{C.~L.} \bibnamefont{Kane}},
  \bibinfo{author}{\bibfnamefont{E.~J.} \bibnamefont{Mele}}, \bibnamefont{and}
  \bibinfo{author}{\bibfnamefont{A.~M.} \bibnamefont{Rappe}},
  \bibinfo{journal}{Phys. Rev. Lett.} \textbf{\bibinfo{volume}{108}},
  \bibinfo{pages}{140405} (\bibinfo{year}{2012}).

\bibitem[{\citenamefont{Wang et~al.}(2012)\citenamefont{Wang, Sun, Chen,
  Franchini, Xu, Weng, Dai, and Fang}}]{Wang2012b}
\bibinfo{author}{\bibfnamefont{Z.}~\bibnamefont{Wang}},
  \bibinfo{author}{\bibfnamefont{Y.}~\bibnamefont{Sun}},
  \bibinfo{author}{\bibfnamefont{X.-Q.} \bibnamefont{Chen}},
  \bibinfo{author}{\bibfnamefont{C.}~\bibnamefont{Franchini}},
  \bibinfo{author}{\bibfnamefont{G.}~\bibnamefont{Xu}},
  \bibinfo{author}{\bibfnamefont{H.}~\bibnamefont{Weng}},
  \bibinfo{author}{\bibfnamefont{X.}~\bibnamefont{Dai}}, \bibnamefont{and}
  \bibinfo{author}{\bibfnamefont{Z.}~\bibnamefont{Fang}},
  \bibinfo{journal}{Phys. Rev. B} \textbf{\bibinfo{volume}{85}},
  \bibinfo{pages}{195320} (\bibinfo{year}{2012}).

\bibitem[{\citenamefont{Wang et~al.}(2013)\citenamefont{Wang, Weng, Wu, Dai,
  and Fang}}]{Wang2013b}
\bibinfo{author}{\bibfnamefont{Z.}~\bibnamefont{Wang}},
  \bibinfo{author}{\bibfnamefont{H.}~\bibnamefont{Weng}},
  \bibinfo{author}{\bibfnamefont{Q.}~\bibnamefont{Wu}},
  \bibinfo{author}{\bibfnamefont{X.}~\bibnamefont{Dai}}, \bibnamefont{and}
  \bibinfo{author}{\bibfnamefont{Z.}~\bibnamefont{Fang}},
  \bibinfo{journal}{Phys. Rev. B} \textbf{\bibinfo{volume}{88}},
  \bibinfo{pages}{125427} (\bibinfo{year}{2013}).

\bibitem[{\citenamefont{Zhao and Wang}(2013)}]{Zhao2013c}
\bibinfo{author}{\bibfnamefont{Y.~X.} \bibnamefont{Zhao}} \bibnamefont{and}
  \bibinfo{author}{\bibfnamefont{Z.~D.} \bibnamefont{Wang}},
  \bibinfo{journal}{Phys. Rev. Lett.} \textbf{\bibinfo{volume}{110}},
  \bibinfo{pages}{240404} (\bibinfo{year}{2013}).

\bibitem[{\citenamefont{Yang and Nagaosa}(2014)}]{Yang2014a}
\bibinfo{author}{\bibfnamefont{B.-J.} \bibnamefont{Yang}} \bibnamefont{and}
  \bibinfo{author}{\bibfnamefont{N.}~\bibnamefont{Nagaosa}},
  \bibinfo{journal}{Nat Commun} \textbf{\bibinfo{volume}{5}},
  \bibinfo{pages}{4898} (\bibinfo{year}{2014}).

\bibitem[{\citenamefont{Weng et~al.}(2015{\natexlab{a}})\citenamefont{Weng,
  Fang, Fang, Bernevig, and Dai}}]{Weng2015}
\bibinfo{author}{\bibfnamefont{H.}~\bibnamefont{Weng}},
  \bibinfo{author}{\bibfnamefont{C.}~\bibnamefont{Fang}},
  \bibinfo{author}{\bibfnamefont{Z.}~\bibnamefont{Fang}},
  \bibinfo{author}{\bibfnamefont{B.~A.} \bibnamefont{Bernevig}},
  \bibnamefont{and} \bibinfo{author}{\bibfnamefont{X.}~\bibnamefont{Dai}},
  \bibinfo{journal}{Phys. Rev. X} \textbf{\bibinfo{volume}{5}},
  \bibinfo{pages}{011029} (\bibinfo{year}{2015}{\natexlab{a}}).

\bibitem[{\citenamefont{Huang et~al.}(2015)\citenamefont{Huang, Xu, Belopolski,
  Lee, Chang, Wang, Alidoust, Bian, Neupane, Zhang et~al.}}]{Huang2015}
\bibinfo{author}{\bibfnamefont{S.-M.} \bibnamefont{Huang}},
  \bibinfo{author}{\bibfnamefont{S.-Y.} \bibnamefont{Xu}},
  \bibinfo{author}{\bibfnamefont{I.}~\bibnamefont{Belopolski}},
  \bibinfo{author}{\bibfnamefont{C.-C.} \bibnamefont{Lee}},
  \bibinfo{author}{\bibfnamefont{G.}~\bibnamefont{Chang}},
  \bibinfo{author}{\bibfnamefont{B.}~\bibnamefont{Wang}},
  \bibinfo{author}{\bibfnamefont{N.}~\bibnamefont{Alidoust}},
  \bibinfo{author}{\bibfnamefont{G.}~\bibnamefont{Bian}},
  \bibinfo{author}{\bibfnamefont{M.}~\bibnamefont{Neupane}},
  \bibinfo{author}{\bibfnamefont{C.}~\bibnamefont{Zhang}},
  \bibnamefont{et~al.}, \bibinfo{journal}{Nat Commun}
  \textbf{\bibinfo{volume}{6}}, \bibinfo{pages}{7373} (\bibinfo{year}{2015}).

\bibitem[{\citenamefont{Liu et~al.}(2014)\citenamefont{Liu, Zhou, Zhang, Wang,
  Weng, Prabhakaran, Mo, Shen, Fang, Dai et~al.}}]{Liu2014c}
\bibinfo{author}{\bibfnamefont{Z.~K.} \bibnamefont{Liu}},
  \bibinfo{author}{\bibfnamefont{B.}~\bibnamefont{Zhou}},
  \bibinfo{author}{\bibfnamefont{Y.}~\bibnamefont{Zhang}},
  \bibinfo{author}{\bibfnamefont{Z.~J.} \bibnamefont{Wang}},
  \bibinfo{author}{\bibfnamefont{H.~M.} \bibnamefont{Weng}},
  \bibinfo{author}{\bibfnamefont{D.}~\bibnamefont{Prabhakaran}},
  \bibinfo{author}{\bibfnamefont{S.-K.} \bibnamefont{Mo}},
  \bibinfo{author}{\bibfnamefont{Z.~X.} \bibnamefont{Shen}},
  \bibinfo{author}{\bibfnamefont{Z.}~\bibnamefont{Fang}},
  \bibinfo{author}{\bibfnamefont{X.}~\bibnamefont{Dai}}, \bibnamefont{et~al.},
  \bibinfo{journal}{Science} \textbf{\bibinfo{volume}{343}},
  \bibinfo{pages}{864} (\bibinfo{year}{2014}).

\bibitem[{\citenamefont{Borisenko et~al.}(2014)\citenamefont{Borisenko, Gibson,
  Evtushinsky, Zabolotnyy, B\"uchner, and Cava}}]{Borisenko2014}
\bibinfo{author}{\bibfnamefont{S.}~\bibnamefont{Borisenko}},
  \bibinfo{author}{\bibfnamefont{Q.}~\bibnamefont{Gibson}},
  \bibinfo{author}{\bibfnamefont{D.}~\bibnamefont{Evtushinsky}},
  \bibinfo{author}{\bibfnamefont{V.}~\bibnamefont{Zabolotnyy}},
  \bibinfo{author}{\bibfnamefont{B.}~\bibnamefont{B\"uchner}},
  \bibnamefont{and} \bibinfo{author}{\bibfnamefont{R.~J.} \bibnamefont{Cava}},
  \bibinfo{journal}{Phys. Rev. Lett.} \textbf{\bibinfo{volume}{113}},
  \bibinfo{pages}{027603} (\bibinfo{year}{2014}).

\bibitem[{\citenamefont{Lv et~al.}(2015)\citenamefont{Lv, Weng, Fu, Wang, Miao,
  Ma, Richard, Huang, Zhao, Chen et~al.}}]{Lv2015}
\bibinfo{author}{\bibfnamefont{B.~Q.} \bibnamefont{Lv}},
  \bibinfo{author}{\bibfnamefont{H.~M.} \bibnamefont{Weng}},
  \bibinfo{author}{\bibfnamefont{B.~B.} \bibnamefont{Fu}},
  \bibinfo{author}{\bibfnamefont{X.~P.} \bibnamefont{Wang}},
  \bibinfo{author}{\bibfnamefont{H.}~\bibnamefont{Miao}},
  \bibinfo{author}{\bibfnamefont{J.}~\bibnamefont{Ma}},
  \bibinfo{author}{\bibfnamefont{P.}~\bibnamefont{Richard}},
  \bibinfo{author}{\bibfnamefont{X.~C.} \bibnamefont{Huang}},
  \bibinfo{author}{\bibfnamefont{L.~X.} \bibnamefont{Zhao}},
  \bibinfo{author}{\bibfnamefont{G.~F.} \bibnamefont{Chen}},
  \bibnamefont{et~al.}, \bibinfo{journal}{Phys. Rev. X}
  \textbf{\bibinfo{volume}{5}}, \bibinfo{pages}{031013} (\bibinfo{year}{2015}).

\bibitem[{\citenamefont{Xu et~al.}(2015)\citenamefont{Xu, Belopolski, Alidoust,
  Neupane, Bian, Zhang, Sankar, Chang, Yuan, Lee et~al.}}]{Xu2015a}
\bibinfo{author}{\bibfnamefont{S.-Y.} \bibnamefont{Xu}},
  \bibinfo{author}{\bibfnamefont{I.}~\bibnamefont{Belopolski}},
  \bibinfo{author}{\bibfnamefont{N.}~\bibnamefont{Alidoust}},
  \bibinfo{author}{\bibfnamefont{M.}~\bibnamefont{Neupane}},
  \bibinfo{author}{\bibfnamefont{G.}~\bibnamefont{Bian}},
  \bibinfo{author}{\bibfnamefont{C.}~\bibnamefont{Zhang}},
  \bibinfo{author}{\bibfnamefont{R.}~\bibnamefont{Sankar}},
  \bibinfo{author}{\bibfnamefont{G.}~\bibnamefont{Chang}},
  \bibinfo{author}{\bibfnamefont{Z.}~\bibnamefont{Yuan}},
  \bibinfo{author}{\bibfnamefont{C.-C.} \bibnamefont{Lee}},
  \bibnamefont{et~al.}, \bibinfo{journal}{Science}
  \textbf{\bibinfo{volume}{349}}, \bibinfo{pages}{613} (\bibinfo{year}{2015}).

\bibitem[{\citenamefont{Weng et~al.}(2015{\natexlab{b}})\citenamefont{Weng,
  Liang, Xu, Yu, Fang, Dai, and Kawazoe}}]{Weng2015c}
\bibinfo{author}{\bibfnamefont{H.}~\bibnamefont{Weng}},
  \bibinfo{author}{\bibfnamefont{Y.}~\bibnamefont{Liang}},
  \bibinfo{author}{\bibfnamefont{Q.}~\bibnamefont{Xu}},
  \bibinfo{author}{\bibfnamefont{R.}~\bibnamefont{Yu}},
  \bibinfo{author}{\bibfnamefont{Z.}~\bibnamefont{Fang}},
  \bibinfo{author}{\bibfnamefont{X.}~\bibnamefont{Dai}}, \bibnamefont{and}
  \bibinfo{author}{\bibfnamefont{Y.}~\bibnamefont{Kawazoe}},
  \bibinfo{journal}{Phys. Rev. B} \textbf{\bibinfo{volume}{92}},
  \bibinfo{pages}{045108} (\bibinfo{year}{2015}{\natexlab{b}}).

\bibitem[{\citenamefont{Yang et~al.}(2014)\citenamefont{Yang, Pan, and
  Zhang}}]{Yang2014c}
\bibinfo{author}{\bibfnamefont{S.~A.} \bibnamefont{Yang}},
  \bibinfo{author}{\bibfnamefont{H.}~\bibnamefont{Pan}}, \bibnamefont{and}
  \bibinfo{author}{\bibfnamefont{F.}~\bibnamefont{Zhang}},
  \bibinfo{journal}{Phys. Rev. Lett.} \textbf{\bibinfo{volume}{113}},
  \bibinfo{pages}{046401} (\bibinfo{year}{2014}).

\bibitem[{\citenamefont{Mullen et~al.}(2015)\citenamefont{Mullen, Uchoa, and
  Glatzhofer}}]{Mullen2015}
\bibinfo{author}{\bibfnamefont{K.}~\bibnamefont{Mullen}},
  \bibinfo{author}{\bibfnamefont{B.}~\bibnamefont{Uchoa}}, \bibnamefont{and}
  \bibinfo{author}{\bibfnamefont{D.~T.} \bibnamefont{Glatzhofer}},
  \bibinfo{journal}{Phys. Rev. Lett.} \textbf{\bibinfo{volume}{115}},
  \bibinfo{pages}{026403} (\bibinfo{year}{2015}).

\bibitem[{\citenamefont{Yu et~al.}(2015)\citenamefont{Yu, Weng, Fang, Dai, and
  Hu}}]{Yu2015}
\bibinfo{author}{\bibfnamefont{R.}~\bibnamefont{Yu}},
  \bibinfo{author}{\bibfnamefont{H.}~\bibnamefont{Weng}},
  \bibinfo{author}{\bibfnamefont{Z.}~\bibnamefont{Fang}},
  \bibinfo{author}{\bibfnamefont{X.}~\bibnamefont{Dai}}, \bibnamefont{and}
  \bibinfo{author}{\bibfnamefont{X.}~\bibnamefont{Hu}}, \bibinfo{journal}{Phys.
  Rev. Lett.} \textbf{\bibinfo{volume}{115}}, \bibinfo{pages}{036807}
  (\bibinfo{year}{2015}).

\bibitem[{\citenamefont{Kim et~al.}(2015)\citenamefont{Kim, Wieder, Kane, and
  Rappe}}]{Kim2015a}
\bibinfo{author}{\bibfnamefont{Y.}~\bibnamefont{Kim}},
  \bibinfo{author}{\bibfnamefont{B.~J.} \bibnamefont{Wieder}},
  \bibinfo{author}{\bibfnamefont{C.~L.} \bibnamefont{Kane}}, \bibnamefont{and}
  \bibinfo{author}{\bibfnamefont{A.~M.} \bibnamefont{Rappe}},
  \bibinfo{journal}{Phys. Rev. Lett.} \textbf{\bibinfo{volume}{115}},
  \bibinfo{pages}{036806} (\bibinfo{year}{2015}).

\bibitem[{\citenamefont{Chen et~al.}(2015)\citenamefont{Chen, Xie, Yang, Pan,
  Zhang, Cohen, and Zhang}}]{Chen2015}
\bibinfo{author}{\bibfnamefont{Y.}~\bibnamefont{Chen}},
  \bibinfo{author}{\bibfnamefont{Y.}~\bibnamefont{Xie}},
  \bibinfo{author}{\bibfnamefont{S.~A.} \bibnamefont{Yang}},
  \bibinfo{author}{\bibfnamefont{H.}~\bibnamefont{Pan}},
  \bibinfo{author}{\bibfnamefont{F.}~\bibnamefont{Zhang}},
  \bibinfo{author}{\bibfnamefont{M.~L.} \bibnamefont{Cohen}}, \bibnamefont{and}
  \bibinfo{author}{\bibfnamefont{S.}~\bibnamefont{Zhang}},
  \bibinfo{journal}{Nano Lett.} \textbf{\bibinfo{volume}{15}},
  \bibinfo{pages}{6974} (\bibinfo{year}{2015}).

\bibitem[{\citenamefont{Xie et~al.}(2015)\citenamefont{Xie, Schoop, Seibel,
  Gibson, Xie, and Cava}}]{Xie2015}
\bibinfo{author}{\bibfnamefont{L.~S.} \bibnamefont{Xie}},
  \bibinfo{author}{\bibfnamefont{L.~M.} \bibnamefont{Schoop}},
  \bibinfo{author}{\bibfnamefont{E.~M.} \bibnamefont{Seibel}},
  \bibinfo{author}{\bibfnamefont{Q.~D.} \bibnamefont{Gibson}},
  \bibinfo{author}{\bibfnamefont{W.}~\bibnamefont{Xie}}, \bibnamefont{and}
  \bibinfo{author}{\bibfnamefont{R.~J.} \bibnamefont{Cava}},
  \bibinfo{journal}{APL Materials} \textbf{\bibinfo{volume}{3}},
  \bibinfo{pages}{083602} (\bibinfo{year}{2015}).

\bibitem[{\citenamefont{Fang et~al.}(2015)\citenamefont{Fang, Chen, Kee, and
  Fu}}]{Fang2015}
\bibinfo{author}{\bibfnamefont{C.}~\bibnamefont{Fang}},
  \bibinfo{author}{\bibfnamefont{Y.}~\bibnamefont{Chen}},
  \bibinfo{author}{\bibfnamefont{H.-Y.} \bibnamefont{Kee}}, \bibnamefont{and}
  \bibinfo{author}{\bibfnamefont{L.}~\bibnamefont{Fu}}, \bibinfo{journal}{Phys.
  Rev. B} \textbf{\bibinfo{volume}{92}}, \bibinfo{pages}{081201}
  (\bibinfo{year}{2015}).

\bibitem[{\citenamefont{Chan et~al.}(2016)\citenamefont{Chan, Chiu, Chou, and
  Schnyder}}]{Chan2016}
\bibinfo{author}{\bibfnamefont{Y.-H.} \bibnamefont{Chan}},
  \bibinfo{author}{\bibfnamefont{C.-K.} \bibnamefont{Chiu}},
  \bibinfo{author}{\bibfnamefont{M.~Y.} \bibnamefont{Chou}}, \bibnamefont{and}
  \bibinfo{author}{\bibfnamefont{A.~P.} \bibnamefont{Schnyder}},
  \bibinfo{journal}{Phys. Rev. B} \textbf{\bibinfo{volume}{93}},
  \bibinfo{pages}{205132} (\bibinfo{year}{2016}).

\bibitem[{\citenamefont{Li et~al.}(2016)\citenamefont{Li, Ma, Cheng, Wang, Li,
  Zhang, Li, and Chen}}]{Li2016}
\bibinfo{author}{\bibfnamefont{R.}~\bibnamefont{Li}},
  \bibinfo{author}{\bibfnamefont{H.}~\bibnamefont{Ma}},
  \bibinfo{author}{\bibfnamefont{X.}~\bibnamefont{Cheng}},
  \bibinfo{author}{\bibfnamefont{S.}~\bibnamefont{Wang}},
  \bibinfo{author}{\bibfnamefont{D.}~\bibnamefont{Li}},
  \bibinfo{author}{\bibfnamefont{Z.}~\bibnamefont{Zhang}},
  \bibinfo{author}{\bibfnamefont{Y.}~\bibnamefont{Li}}, \bibnamefont{and}
  \bibinfo{author}{\bibfnamefont{X.-Q.} \bibnamefont{Chen}},
  \bibinfo{journal}{Phys. Rev. Lett.} \textbf{\bibinfo{volume}{117}},
  \bibinfo{pages}{096401} (\bibinfo{year}{2016}).

\bibitem[{\citenamefont{Bian et~al.}(2016)\citenamefont{Bian, Chang, Sankar,
  Xu, Zheng, Neupert, Chiu, Huang, Chang, Belopolski et~al.}}]{Bian2016}
\bibinfo{author}{\bibfnamefont{G.}~\bibnamefont{Bian}},
  \bibinfo{author}{\bibfnamefont{T.-R.} \bibnamefont{Chang}},
  \bibinfo{author}{\bibfnamefont{R.}~\bibnamefont{Sankar}},
  \bibinfo{author}{\bibfnamefont{S.-Y.} \bibnamefont{Xu}},
  \bibinfo{author}{\bibfnamefont{H.}~\bibnamefont{Zheng}},
  \bibinfo{author}{\bibfnamefont{T.}~\bibnamefont{Neupert}},
  \bibinfo{author}{\bibfnamefont{C.-K.} \bibnamefont{Chiu}},
  \bibinfo{author}{\bibfnamefont{S.-M.} \bibnamefont{Huang}},
  \bibinfo{author}{\bibfnamefont{G.}~\bibnamefont{Chang}},
  \bibinfo{author}{\bibfnamefont{I.}~\bibnamefont{Belopolski}},
  \bibnamefont{et~al.}, \bibinfo{journal}{Nature Communications}
  \textbf{\bibinfo{volume}{7}}, \bibinfo{pages}{10556} (\bibinfo{year}{2016}).

\bibitem[{\citenamefont{Schoop et~al.}(2016)\citenamefont{Schoop, Ali,
  Straßer, Topp, Varykhalov, Marchenko, Duppel, Parkin, Lotsch, and
  Ast}}]{Schoop2016}
\bibinfo{author}{\bibfnamefont{L.~M.} \bibnamefont{Schoop}},
  \bibinfo{author}{\bibfnamefont{M.~N.} \bibnamefont{Ali}},
  \bibinfo{author}{\bibfnamefont{C.}~\bibnamefont{Straßer}},
  \bibinfo{author}{\bibfnamefont{A.}~\bibnamefont{Topp}},
  \bibinfo{author}{\bibfnamefont{A.}~\bibnamefont{Varykhalov}},
  \bibinfo{author}{\bibfnamefont{D.}~\bibnamefont{Marchenko}},
  \bibinfo{author}{\bibfnamefont{V.}~\bibnamefont{Duppel}},
  \bibinfo{author}{\bibfnamefont{S.~S.~P.} \bibnamefont{Parkin}},
  \bibinfo{author}{\bibfnamefont{B.~V.} \bibnamefont{Lotsch}},
  \bibnamefont{and} \bibinfo{author}{\bibfnamefont{C.~R.} \bibnamefont{Ast}},
  \bibinfo{journal}{Nature Communications} \textbf{\bibinfo{volume}{7}},
  \bibinfo{pages}{11696} (\bibinfo{year}{2016}).

\bibitem[{HXu()}]{HXu}
\bibinfo{note}{Y. Jin, R. Wang, J. Zhao, C. Zheng, L.-Y. Gan, J. Liu, H. Xu,
  and S. Y. Tong, arXiv:1608.05791.}

\bibitem[{Yu2()}]{Yu2017}
\bibinfo{note}{R. Yu, Q. Wu, Z. Fang, and H. Weng, arXiv:1701.08502.}

\bibitem[{\citenamefont{Parameswaran et~al.}(2013)\citenamefont{Parameswaran,
  Turner, Arovas, and Vishwanath}}]{Parameswaran2013}
\bibinfo{author}{\bibfnamefont{S.~A.} \bibnamefont{Parameswaran}},
  \bibinfo{author}{\bibfnamefont{A.~M.} \bibnamefont{Turner}},
  \bibinfo{author}{\bibfnamefont{D.~P.} \bibnamefont{Arovas}},
  \bibnamefont{and}
  \bibinfo{author}{\bibfnamefont{A.}~\bibnamefont{Vishwanath}},
  \bibinfo{journal}{Nat Phys} \textbf{\bibinfo{volume}{9}},
  \bibinfo{pages}{299} (\bibinfo{year}{2013}).

\bibitem[{\citenamefont{Steinberg et~al.}(2014)\citenamefont{Steinberg, Young,
  Zaheer, Kane, Mele, and Rappe}}]{Steinberg2014}
\bibinfo{author}{\bibfnamefont{J.~A.} \bibnamefont{Steinberg}},
  \bibinfo{author}{\bibfnamefont{S.~M.} \bibnamefont{Young}},
  \bibinfo{author}{\bibfnamefont{S.}~\bibnamefont{Zaheer}},
  \bibinfo{author}{\bibfnamefont{C.~L.} \bibnamefont{Kane}},
  \bibinfo{author}{\bibfnamefont{E.~J.} \bibnamefont{Mele}}, \bibnamefont{and}
  \bibinfo{author}{\bibfnamefont{A.~M.} \bibnamefont{Rappe}},
  \bibinfo{journal}{Phys. Rev. Lett.} \textbf{\bibinfo{volume}{112}},
  \bibinfo{pages}{036403} (\bibinfo{year}{2014}).

\bibitem[{\citenamefont{Young and Kane}(2015)}]{Young2015a}
\bibinfo{author}{\bibfnamefont{S.~M.} \bibnamefont{Young}} \bibnamefont{and}
  \bibinfo{author}{\bibfnamefont{C.~L.} \bibnamefont{Kane}},
  \bibinfo{journal}{Phys. Rev. Lett.} \textbf{\bibinfo{volume}{115}},
  \bibinfo{pages}{126803} (\bibinfo{year}{2015}).

\bibitem[{\citenamefont{Watanabe et~al.}(2016)\citenamefont{Watanabe, Po,
  Zaletel, and Vishwanath}}]{Watanabe2016}
\bibinfo{author}{\bibfnamefont{H.}~\bibnamefont{Watanabe}},
  \bibinfo{author}{\bibfnamefont{H.~C.} \bibnamefont{Po}},
  \bibinfo{author}{\bibfnamefont{M.~P.} \bibnamefont{Zaletel}},
  \bibnamefont{and}
  \bibinfo{author}{\bibfnamefont{A.}~\bibnamefont{Vishwanath}},
  \bibinfo{journal}{Phys. Rev. Lett.} \textbf{\bibinfo{volume}{117}},
  \bibinfo{pages}{096404} (\bibinfo{year}{2016}).

\bibitem[{\citenamefont{Bradlyn et~al.}(2016)\citenamefont{Bradlyn, Cano, Wang,
  Vergniory, Felser, Cava, and Bernevig}}]{Bradlyn2016}
\bibinfo{author}{\bibfnamefont{B.}~\bibnamefont{Bradlyn}},
  \bibinfo{author}{\bibfnamefont{J.}~\bibnamefont{Cano}},
  \bibinfo{author}{\bibfnamefont{Z.}~\bibnamefont{Wang}},
  \bibinfo{author}{\bibfnamefont{M.~G.} \bibnamefont{Vergniory}},
  \bibinfo{author}{\bibfnamefont{C.}~\bibnamefont{Felser}},
  \bibinfo{author}{\bibfnamefont{R.~J.} \bibnamefont{Cava}}, \bibnamefont{and}
  \bibinfo{author}{\bibfnamefont{B.~A.} \bibnamefont{Bernevig}},
  \bibinfo{journal}{Science} p. \bibinfo{pages}{10.1126/science.aaf5037}
  (\bibinfo{year}{2016}).

\bibitem[{\citenamefont{Liang et~al.}(2016)\citenamefont{Liang, Zhou, Yu, Wang,
  and Weng}}]{Liang2016}
\bibinfo{author}{\bibfnamefont{Q.-F.} \bibnamefont{Liang}},
  \bibinfo{author}{\bibfnamefont{J.}~\bibnamefont{Zhou}},
  \bibinfo{author}{\bibfnamefont{R.}~\bibnamefont{Yu}},
  \bibinfo{author}{\bibfnamefont{Z.}~\bibnamefont{Wang}}, \bibnamefont{and}
  \bibinfo{author}{\bibfnamefont{H.}~\bibnamefont{Weng}},
  \bibinfo{journal}{Phys. Rev. B} \textbf{\bibinfo{volume}{93}},
  \bibinfo{pages}{085427} (\bibinfo{year}{2016}).

\bibitem[{\citenamefont{Wieder and Kane}(2016)}]{Wieder2016}
\bibinfo{author}{\bibfnamefont{B.~J.} \bibnamefont{Wieder}} \bibnamefont{and}
  \bibinfo{author}{\bibfnamefont{C.~L.} \bibnamefont{Kane}},
  \bibinfo{journal}{Phys. Rev. B} \textbf{\bibinfo{volume}{94}},
  \bibinfo{pages}{155108} (\bibinfo{year}{2016}).

\bibitem[{\citenamefont{Zhao and Schnyder}(2016)}]{Zhao2016a}
\bibinfo{author}{\bibfnamefont{Y.~X.} \bibnamefont{Zhao}} \bibnamefont{and}
  \bibinfo{author}{\bibfnamefont{A.~P.} \bibnamefont{Schnyder}},
  \bibinfo{journal}{Phys. Rev. B} \textbf{\bibinfo{volume}{94}},
  \bibinfo{pages}{195109} (\bibinfo{year}{2016}).

\bibitem[{RCh()}]{RChen}
\bibinfo{note}{R. Chen, H. C. Po, J. B. Neaton, and A. Vishwanath,
  arXiv:1611.06860.}

\bibitem[{\citenamefont{Fang et~al.}(2016)\citenamefont{Fang, Weng, Dai, and
  Fang}}]{Fang2016}
\bibinfo{author}{\bibfnamefont{C.}~\bibnamefont{Fang}},
  \bibinfo{author}{\bibfnamefont{H.}~\bibnamefont{Weng}},
  \bibinfo{author}{\bibfnamefont{X.}~\bibnamefont{Dai}}, \bibnamefont{and}
  \bibinfo{author}{\bibfnamefont{Z.}~\bibnamefont{Fang}},
  \bibinfo{journal}{Chinese Physics B} \textbf{\bibinfo{volume}{25}},
  \bibinfo{pages}{117106} (\bibinfo{year}{2016}).

\bibitem[{\citenamefont{Chen et~al.}(2016)\citenamefont{Chen, Kim, and
  Kee}}]{Chen2016b}
\bibinfo{author}{\bibfnamefont{Y.}~\bibnamefont{Chen}},
  \bibinfo{author}{\bibfnamefont{H.-S.} \bibnamefont{Kim}}, \bibnamefont{and}
  \bibinfo{author}{\bibfnamefont{H.-Y.} \bibnamefont{Kee}},
  \bibinfo{journal}{Phys. Rev. B} \textbf{\bibinfo{volume}{93}},
  \bibinfo{pages}{155140} (\bibinfo{year}{2016}).

\bibitem[{\citenamefont{Yang et~al.}(2017)\citenamefont{Yang, Bojesen,
  Morimoto, and Furusaki}}]{Yang2017}
\bibinfo{author}{\bibfnamefont{B.-J.} \bibnamefont{Yang}},
  \bibinfo{author}{\bibfnamefont{T.~A.} \bibnamefont{Bojesen}},
  \bibinfo{author}{\bibfnamefont{T.}~\bibnamefont{Morimoto}}, \bibnamefont{and}
  \bibinfo{author}{\bibfnamefont{A.}~\bibnamefont{Furusaki}},
  \bibinfo{journal}{Phys. Rev. B} \textbf{\bibinfo{volume}{95}},
  \bibinfo{pages}{075135} (\bibinfo{year}{2017}).

\bibitem[{Fur()}]{Furusaki}
\bibinfo{note}{A. Furusaki, arXiv:1702.07606.}

\bibitem[{Tak()}]{Takahashi}
\bibinfo{note}{R. Takahashi, M. Hirayama, and S. Murakami, arXiv:1704.02151.}

\bibitem[{\citenamefont{Wang et~al.}(2016)\citenamefont{Wang, Alexandradinata,
  Cava, and Bernevig}}]{Wang2016a}
\bibinfo{author}{\bibfnamefont{Z.}~\bibnamefont{Wang}},
  \bibinfo{author}{\bibfnamefont{A.}~\bibnamefont{Alexandradinata}},
  \bibinfo{author}{\bibfnamefont{R.~J.} \bibnamefont{Cava}}, \bibnamefont{and}
  \bibinfo{author}{\bibfnamefont{B.~A.} \bibnamefont{Bernevig}},
  \bibinfo{journal}{Nature} \textbf{\bibinfo{volume}{532}},
  \bibinfo{pages}{189} (\bibinfo{year}{2016}).

\bibitem[{Ma2()}]{Ma2015}
\bibinfo{note}{J.-Z. Ma, C.-J. Yi, B. Q. Lv, Z. J. Wang, S.-M. Nie, L. Wang,
  L.-Y. Kong, Y.-B. Huang, P. Richard, H.-M. Weng, B. A. Bernevig, Y.-G. Shi,
  T. Qian, and H. Ding, arXiv:1605.06824.}

\bibitem[{Wan()}]{Wang2017}
\bibinfo{note}{L. Wang, S.-K. Jian, and H. Yao, arXiv:1702.06140.}

\bibitem[{\citenamefont{Bzdušek et~al.}(2016)\citenamefont{Bzdušek, Wu,
  Rüegg, Sigrist, and Soluyanov}}]{Bzdusek2016}
\bibinfo{author}{\bibfnamefont{T.}~\bibnamefont{Bzdušek}},
  \bibinfo{author}{\bibfnamefont{Q.}~\bibnamefont{Wu}},
  \bibinfo{author}{\bibfnamefont{A.}~\bibnamefont{Rüegg}},
  \bibinfo{author}{\bibfnamefont{M.}~\bibnamefont{Sigrist}}, \bibnamefont{and}
  \bibinfo{author}{\bibfnamefont{A.~A.} \bibnamefont{Soluyanov}},
  \bibinfo{journal}{Nature} \textbf{\bibinfo{volume}{538}}, \bibinfo{pages}{75}
  (\bibinfo{year}{2016}).

\bibitem[{\citenamefont{Magn\'{e}li}(1957)}]{Magneli1957}
\bibinfo{author}{\bibfnamefont{A.}~\bibnamefont{Magn\'{e}li}},
  \bibinfo{journal}{Acta Chem. Scand.} \textbf{\bibinfo{volume}{11}},
  \bibinfo{pages}{28} (\bibinfo{year}{1957}).

\bibitem[{\citenamefont{Colaitis and Lécaille}(1972)}]{Colaitis1972}
\bibinfo{author}{\bibfnamefont{D.}~\bibnamefont{Colaitis}} \bibnamefont{and}
  \bibinfo{author}{\bibfnamefont{C.}~\bibnamefont{Lécaille}},
  \bibinfo{journal}{Materials Research Bulletin} \textbf{\bibinfo{volume}{7}},
  \bibinfo{pages}{369} (\bibinfo{year}{1972}).

\bibitem[{\citenamefont{Goodenough et~al.}(1965)\citenamefont{Goodenough,
  Gibart, and Brenet}}]{Goodenough1965}
\bibinfo{author}{\bibfnamefont{J.~B.} \bibnamefont{Goodenough}},
  \bibinfo{author}{\bibfnamefont{P.}~\bibnamefont{Gibart}}, \bibnamefont{and}
  \bibinfo{author}{\bibfnamefont{J.}~\bibnamefont{Brenet}},
  \bibinfo{journal}{CR Hebd. S\'{e}ances Acad. Sci.}
  \textbf{\bibinfo{volume}{261}}, \bibinfo{pages}{2331} (\bibinfo{year}{1965}).

\bibitem[{sup()}]{supp}
\bibinfo{note}{See Supplemental Material.}

\bibitem[{\citenamefont{Heikkil\"{a} et~al.}(2011)\citenamefont{Heikkil\"{a},
  Kopnin, and Volovik}}]{Heikkilae2011}
\bibinfo{author}{\bibfnamefont{T.~T.} \bibnamefont{Heikkil\"{a}}},
  \bibinfo{author}{\bibfnamefont{N.~B.} \bibnamefont{Kopnin}},
  \bibnamefont{and} \bibinfo{author}{\bibfnamefont{G.~E.}
  \bibnamefont{Volovik}}, \bibinfo{journal}{JETP Letters}
  \textbf{\bibinfo{volume}{94}}, \bibinfo{pages}{233} (\bibinfo{year}{2011}).

\end{thebibliography}


\end{document}